\documentstyle[epsfig]{jaa}

\DeclareMathAlphabet{\mathsc}{OT1}{cmr}{m}{sc}
\def\testbx{bx}%
\DeclareRobustCommand{\ion}[2]{%
\relax\ifmmode
\ifx\testbx\f@series
{\mathbf{#1\,\mathsc{#2}}}\else
{\mathrm{#1\,\mathsc{#2}}}\fi
\else\textup{#1\,{\mdseries\textsc{#2}}}%
\fi}

\begin{document}
\title[Radio halos in galaxy clusters]
{Giant radio halos in galaxy clusters as probes of particle acceleration 
in turbulent regions}
\author[G. Brunetti] 
{G. Brunetti
\thanks{INAF- Istituto di Radioastronomia, via P.Gobetti 101, I--40129,
Bologna, Italy}
\thanks{e-mail: brunetti@ira.inaf.it}}

\pubyear{xxxx}
\volume{xx}
\date{Received xxx; accepted xxx}
\maketitle
\label{firstpage}
\begin{abstract}

\end{abstract}
Giant radio halos in galaxy clusters probe mechanisms of particle
acceleration connected with cluster merger events.
Shocks and turbulence are driven in the inter-galactic-medium
(IGM) during clusters mergers and may have a deep impact on the non-thermal
properties of galaxy clusters.
Models of turbulent (re)acceleration of relativistic particles 
allow good correspondence with present observations, 
from radio halos to $\gamma$-ray upper limits, although 
several aspects of this complex 
scenario remain still poorly understood.

After providing basic motivations for turbulent acceleration
in galaxy clusters, we discuss relevant aspects of
the physics of particle acceleration
by MHD turbulence and the expected broad--band non-thermal emission from 
galaxy clusters.
We discuss (in brief) the most important results of turbulent
(re)acceleration models, the open problems, and the 
possibilities to test models with future observations.
In this respect, further constraints on the origin of giant 
nearby radio halos can also be obtained by combining their (spectral and
morphological) properties with the constraints 
from $\gamma$-ray observations of their parent clusters.

\section{Introduction}

Radio observations show the presence of diffuse (on Mpc scale)
radio emission in a fraction of massive galaxy clusters,
{\it radio halos} from cluster X-ray emitting regions,
and {\it relics}, typically
in the clusters peripheral regions (e.g., Ferrari et al
2008, Venturi 2011 for recent reviews).
Giant radio halos are the most spectacular and best studied 
cluster-scale non-thermal sources. They probe the existence of
complex mechanisms, responsible for their origin, that are 
still poorly understood.

\noindent
Several sources of relativistic particles exist in 
galaxy clusters : ordinary and active galaxies (AGN), 
and cosmological shock waves (e.g. Blasi et al 2007 for a review).
However the time necessary to GeV electrons\footnote{
those responsible for the synchrotron radiation in the radio band} 
to diffuse on Mpc (halo) scales from these sources is much longer than their
radiative life-time ($\sim 10^8$ yrs).
Thus radio halos prove processes of acceleration/injection of
GeV electrons that must be ``distributed'' on cluster scales (Jaffe 1977).

Giant radio halos are not common in galaxy clusters and observed only
in about $1/3$ of the most massive systems (e.g. Giovannini et al 1999,
Kempner \& Sarazin 2001, Cassano et al. 2008).
Radio observations and their follow up in the X-rays suggested 
that radio halos are found only in dynamically disturbed systems 
(e.g. Buote 2001, Govoni et al 2004).
More recently, the sensitivity of the Radio Halo Survey at 
the GMRT (Venturi et al 2007, 08) allows for starting
a solid statistical exploration of
clusters radio properties. It allows the discovery of the
clusters radio bimodality that pin-points 
the {\it transient} nature of radio halos that are generated 
in connection with clusters mergers and fade away
when clusters become more relaxed systems
(e.g. Brunetti et al. 2009, Cassano et al 2010a).

These observations suggest that a fraction of the gravitational energy 
that is dissipated during merger events is channelled into the acceleration 
of non-thermal components.
In this case the scenario for the origin of radio halos 
assumes that relativistic particles are (re)accelerated in Mpc regions
by MHD turbulence generated during cluster mergers (e.g., Brunetti et al 2001,
Petrosian 2001), 
this may naturally explain the tight connection between
halos and mergers. Alternative possibilities that have been proposed
so far for the origin of the emitting electrons
include the generation of secondary electrons due to proton-proton
collisions in the IGM (e.g. Blasi \& Colafrancesco 1999, Pfrommer \&
En\ss lin 2004, Keshet \& Loeb 2010), and dark-matter annihilation 
in the cluster volume (e.g. Colafrancesco, Profumo, Ullio 2006).
Here we discuss the case of the turbulent (re)acceleration scenario.

\section{Turbulence and turbulent acceleration in galaxy clusters}

\subsection{Why turbulent acceleration ? -- A simple motivation}

Observations constrain models of giant
radio halos, in several cases putting some tension on a
``pure'' secondary origin of the emitting electrons (e.g. Ferrari et
al.~2008 for review; Brunetti et al. 2008, 09, Donnert et al. 2010a,b,
Jeltema \& Profumo 2011, Brown \& Rudnick 2011,
Bonafede et al. 2011 for recent results).

\noindent
In this Section we focus on the observed spectral properties 
of giant radio halos
that provide part of the motivation for turbulent acceleration 
of the emitting electrons.
Potentially the synchrotron spectrum gives 
information on the efficiency of 
the acceleration of the emitting electrons.
The maximum energy of electrons is given by the competition between
acceleration efficieny and (radiative) losses, $E_{max} \approx
\chi(E)/\beta_{rad}$ ($\chi(E) = \chi$ for FERMI mechanisms,
and $\beta_{rad} = c_2 (B^2 + B_{IC}^2)\,$).
Consequently the maximum frequency of the synchrotron radiation from 
the accelerated electrons (at higher frequencies the spectrum 
steepens), $\nu_{max} = c_1 B E_{max}^2$, is :

\begin{equation}
\nu_{max} 
\sim {{c_1}\over{c_2^2}}
{{ B \chi^2}\over{(B^2 + B_{IC}^2)^2}}
\end{equation}

\noindent
Assuming that inverse Compton (IC) 
and synchrotron losses are of the same order of
magnitude, i.e. $B \approx {\rm few}\, \mu$G 
as suggested by the analysis
of Rotation Measures (RM) of cluster radio sources
(e.g., Bonafede et al.~2011 and ref. therein), the measure 
of $\nu_{max}$ allows for estimating $\chi$ and the acceleration
time-scale $\tau_{acc}\sim 1/\chi$.

The ``hystorical'' motivation for turbulent acceleration for the origin
of radio halos comes from the 
spectrum of the Coma halo, the prototype 
of these sources (e.g., Willson 1970, Giovannini et al 1993).
Coma is the unique halo with a spectrum measured
over a wide frequency range (Fig.~1a).
The spectrum significantly steepens at higher frequencies : 
a power-law that fits the data at lower
frequencies overestimates the flux measured at 2.7 and 5 GHz
by a factor 2 and 3, respectively (e.g. Thierbach et al 2003)\footnote{
even by considering the effect of the SZ-decrement (see also
Donnert et al.(2010a)}.
The observed steepening of Coma implies 
(from Eq.~1) $\tau_{acc} \approx 10^8$yrs, i.e. that ``gentle'' 
(poorly efficient) and spatially-distributed 
(on Mpc scales) mechanisms must be responsible for the acceleration of the 
emitting electrons; the most natural candidate 
is acceleration by turbulence, that is indeed poorly
efficient (e.g. Schlickeiser et al. 1987).

Also the spectrum  of others radio halos favours turbulent acceleration. 
Although the spectrum of giant radio halos is still poorly known, 
and less than 10-12 halos are observed at 2 frequencies, 
the observed values of the spectral indices span
a broad range, $\alpha \sim 1 - 2$ 
\footnote{The upper bound of the range 
is probably limited by the fact that steeper halos
would be difficult to observe with present radio telescopes}
($F(\nu)\propto \nu^{-\alpha}$,
e.g. Venturi 2011).
This readily implies that the synchrotron spectrum
of radio halos is far from being
a ``universal'' power law and {\it poses crucial 
constraints to the nature of the mechanisms that generate these
sources}.
In particular, halos with extreme spectral properties, $\alpha \sim 1.5-2$
(e.g., Brunetti et al 2008, Brentjens 2008, Giovannini et al 2009, 
Macario et al 2010) are important.
Energy arguments rule out the possibility that they have a
(very steep) power law spectrum extending to lower frequencies
and also allow to disfavour 
a ``pure'' secondary origin of the emitting electrons (Brunetti et al 2008,
Dallacasa et al 2009).
Giant radio halos with $\alpha > 1.5$ are explained by assuming 
that their spectrum starts steepening at lower frequencies.
It implies that present observations ``just'' sample 
the range of frequencies where the steepening becomes severe.
According to turbulent acceleration models, 
these very steep-spectrum sources are the halos 
generated with the smaller acceleration efficiency ($\tau_{acc} \approx
2-3 \times 10^8$yrs) among the presently observed radio halos. 

\begin{figure}
\centering
\includegraphics[width=12.5cm, angle=0]{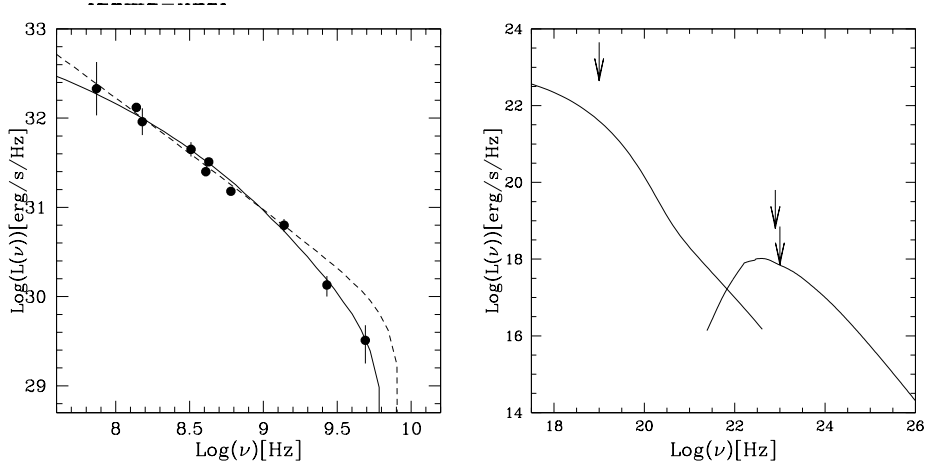}
\caption{\footnotesize
Radio (left) and high-energy (right) spectrum of Coma.
Models, {\it power law} spectrum (dashed) and {\it reacceleration} (solid),
are shown including the effect (cut-off) due to SZ-decrement at high radio
frequencies. Model details are in Brunetti \& Lazarian (2011b), 
relevant data are in Ackermann et
al.(2010), Deiss et al. (1997) and Pizzo (2010).}
\label{fig:size_dens}
\end{figure}

\subsection{Turbulence in galaxy clusters}

Cosmological numerical simulations show that large-scale turbulent motions
are generated during the process of cluster formation 
(e.g., Dolag et al. 2005, Iapichino \& Niemeyer 2008, Vazza et al. 2011).
These motions, with typical velocities $V_L \sim 500-700$ km/s, are
injected at large scales, $L_o \sim 300-500$ kpc, during merging
events and may provide 
the driver for turbulence at smaller scales.

\noindent
Theoretically the 
viscosity in a turbulent and magnetised IGM is strongly suppressed due
to the effect of the bending of magnetic field lines and of the
perturbations of the magnetic field induced by plasma 
instabilities (Schekochihin et al. 2005, Lazarian 2006).
Consequently an inertial range in the IGM
may be established down to collisionless scales where a fraction of the 
turbulent energy is channelled into acceleration/heating of cosmic
rays and thermal plasma. 
At this point we may tought of several processes that can channel
(at least a fraction of)
the turbulent-energy into the (re)acceleration of particles.
They include resonant and non-resonant 
couplings and their efficiency depends on the properties of
turbulence and of the background magnetised
plasma (e.g. Cho \& Lazarian 2006 for review).

\subsection{Turbulent acceleration models for the origin of giant
radio halos and consequences for high energy emission from
galaxy clusters}

Acceleration of electrons from the thermal pool to relativistic energies
by MHD turbulence in the IGM faces serious problems due to energy arguments
(e.g. Petrosian \& East 2008).
Consequently, turbulent acceleration models must assume
a pre-existing population of relativistic particles that
provides the seeds to ``reaccelerate'' during cluster mergers (e.g.
Brunetti 2003, Petrosian \& Bykov 2008 for reviews).

\noindent
To account for the turbulence--particles interaction properly, one must know
both the scaling of turbulence,
the changes with time of turbulence spectrum due to
the most relevant damping processes,
and the interactions of turbulence with various waves produced by cosmic
rays. For this reason the fraction of the turbulent energy that gets into
(re)acceleration of cosmic rays in the IGM is uncertain and reflects 
our ignorance of the details of the properties of turbulence and
of the (connected) micro-physics of the IGM.

Cases where a large fraction of the turbulent energy is dissipated 
into the (re)acceleration of cosmic rays in galaxy clusters include
the gyro-resonant interaction with Alfv\'en modes (e.g. 
Ohno et al 2002, Fujita et al. 2003,
Brunetti et al. 2004)\footnote{In this case it must be
postulated an injection of Alfv\'en modes at quasi--resonant (small)
scales to have quasi--isotropic distribution of the modes (see
Yan \& Lazarian 2004)} and the
resonant (mainly Transit-Time-Damping) interaction with fast modes 
under the assumption that the collisionless scale of the IGM is much
smaller than the Coulomb ion mean free path (e.g. Brunetti \& Lazarian 
2011a)\footnote{In this case it is proposed that the perturbations of 
the magnetic field generated by turbulence-driven
plasma instabilities 
reduce the effective mean free path}.
In these cases the efficiency of particle acceleration is self-regulated
by the back--reaction (damping) of particles on the spectrum 
of turbulence (Brunetti et al. 2004, Brunetti \& Lazarian 2011a).
Stronger turbulence induces more efficient acceleration 
leading to a faster growth of the energy density of cosmic rays with time.
This -- however -- increases the damping of turbulence 
and the interaction approaches a quasi--asymptotic (and very comples)
regime where
cosmic rays get in (quasi) equipartition with turbulence and self-regulate
their (re)acceleration.

According to a more standard approach, the damping of turbulence in the
IGM is dominated by the interaction with the hot IGM.
In this case it is calculated that
only a fraction ($\sim$ 10\%) of turbulence 
goes into the (re)acceleration of cosmic rays (e.g. Cassano \& Brunetti 2005, 
Brunetti \& Lazarian 2007).
This scenario 
is motivated (i) by the idea that the compressible part
of the MHD turbulence contributes the most of the particle 
acceleration in the IGM and (ii) by the fact that fast modes are 
strongly damped (via Transit--Time--Damping) in a hot (and high beta) 
plasma such as the IGM.
This scenario allows prompt calculations of particle acceleration by
MHD turbulence in the IGM.
In Brunetti \& Lazarian (2007) we considered the advances in the
theory of MHD turbulence to develop a comprehensive picture of
turbulence in the IGM and to study the reacceleration of relativistic
particles considering all the
relevant damping processes. 
We have shown that the ensuing cluster-scale radio emission generated 
in merging clusters is in very good agreement with present 
observations of radio halos.

\noindent
More recently we extended our investigation to the case of the 
(re)acceleration of cosmic ray protons and of the secondary electrons
generated in the IGM via pp collisions (Brunetti \& Lazarian 2011b).
These calculations were motivated by the fact that cosmic ray protons 
are long--living particles that are confined (and accumulated) in
clusters (V\"olk et al 1996, Berezinsky et al 1997).
The consequence is the unavoideable generation of secondary particles and
$\gamma$-rays (at some level) in the IGM (e.g., Blasi \& Colafrancesco 1999,
Miniati 2003, Pfrommer \& En\ss lin 2004).
Calculations in Brunetti \& Lazarian (2011b) allow a self-consistent
treatment of the interaction between compressible turbulence and
cosmic rays in the IGM and a complete modeling of the non-thermal
spectrum from galaxy clusters. 
Figure 1 shows the expected spectrum in the case of a Coma--like
cluster where the energy content of compressible 
turbulence and cosmic ray protons is assumed $\approx 18\%$ and 
few $\%$ of that of the IGM, respectively.
The spectrum in Fig.~1 is 
calculated by assuming the magnetic field in the Coma cluster 
(strength and radial profile) as derived from RM (Bonafede
et al.~2010). Under these conditions the expected $\gamma$-ray emission is
about 5--7 times below the upper limits from the first 18 months of 
observations with the FERMI satellite.
A detection with FERMI after $\sim$2 yrs of observations 
would be reconciled by ``postulating'' a magnetic field $\approx 2.5$ times 
smaller than that from RM.

\section{Halo--merger connection and future observations}

The formation and evolution of radio halos depend on the dynamics
of the hosting clusters (see Brunetti et al 2009 for a more detailed 
discussion). 
Observations prove this tight connection, namely
that all radio halos are observed in dynamically disturbed
systems (e.g. Govoni et al 2004, Cassano et al 2010a; see Cassano this
conference).
Merger--turbulence decays at smaller scales
in about one eddy turnover time, few $10^8$yrs, implying a 
temporal connection between mergers (the duration of cluster-cluster
interaction is $>$ Gyr), turbulent (re)acceleration and
radio halos (the particle acceleration time--scale required
for the acceleration of radio emitting electrons is $10^8$yrs). 
Compressible turbulence dissipates most
of its energy in a few eddy turnover times, as soon as galaxy clusters 
becomes more relaxed. 
It implies that radio halos must fade away in more relaxed
systems in a relatively short ($<$ Gyr) time.
Obviously the situation becomes more complex thinking 
of the process of cluster formation that would generate a more complex evolution
of cluster turbulence (e.g. Paul et al 2011,
Vazza et al 2011). Future cosmological simulations that include
a proper treatment of cosmic ray acceleration/cooling will shead light
on this connection.

\begin{figure}
\centering
\includegraphics[width=12.5cm, angle=0]{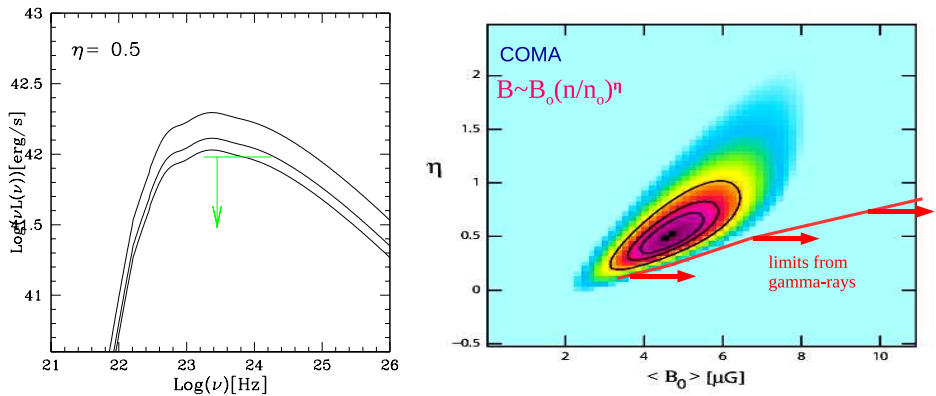}
\caption{\footnotesize
(left) $\gamma$-ray emission ($\pi^o$ decay) from the Coma cluster assuming
secondary models ``forced'' to reproduce the radio brightness and 
luminosity of the halo. Calculations assume $\eta$=0.5 and
$B_o$=4.7, 6.4, 7.1 $\mu$G (from top to bottom).
(right) Comparison between the minimum $B_o$ from $\gamma$-ray upper limits
(assuming secondaries) and the magnetic field derived from RM (contours
refer to 1, 2, 3 $\sigma$ conf. level, from Bonafede et al.~2010).}
\label{fig:VLA1}
\end{figure}

\noindent
A different (yet connected) point is whether all merging clusters should host
giant radio halos.
Observations show several cases of merging
clusters that do not host detectable radio halos (e.g., Cassano et al 2010a,
Russell et al. 2011). From a ``naive'' point of view these systems could be
very young mergers where the decay of turbulence at smaller scales
and the acceleration of particles are not started yet.

\noindent
We may suggest
a more physical explenation in the context of turbulent models.
The frequency where a steepening is predicted in the spectra of radio halos,
$\nu_{max}$ (Sect.~2.1), is determined 
by the fraction of turbulent energy converted into electron re-acceleration.
Only the most energetic merger-events in the Universe can generate
giant radio halos with $\nu_{max} \geq$ 1 GHz (Cassano \& Brunetti 2005).
The generation of these radio halos in less massive 
systems ($M_v \leq 1-2\times
10^{15}\,M_{\odot}$) or in clusters at higher redshifts ($z \geq 0.4-0.5$) is
rare and we expect that 
halos in these systems are mainly generated with
their spectra steepening at lower 
frequencies and are difficult to observe at higher frequencies
(Cassano, Brunetti, Setti 2006).
Interestingly most disturbed systems without observable radio
halos are clusters with $L_X \leq 7-8\times 10^{44}$erg/s (Cassano et al
2010a, Russell et al 2011), thus we might guess that 
these systems have halos that glow up when observed 
at lower radio frequencies (e.g. with LOFAR, LWA).

\noindent
These radio halos with very steep spectrum are predicted to be more frequent
in galaxy clusters, since they can be generated
in connection with less energetic mergers, e.g. between
less massive systems or minor mergers in massive systems, that are more
common in the Universe.
{\it The existence of these radio halos is the most important 
expectation of turbulent models and it stems from the fact
that turbulent acceleration is a poorly efficient process}
(e.g. Brunetti et al 2008).
Crucial tests will come from the future surveys with LOFAR
(Cassano et al 2010b, Rottgering et al 2010).

\section{$\gamma$-rays from galaxy clusters and origin of radio halos}

The confinement of cosmic ray protons in galaxy clusters leads to the
important expectation that clusters must be $\gamma$-ray
emitters due to the production of secondary particles
(V\"olk et al 1996, Berezynsky et al 1997).
The ratio of $\gamma$-rays ($\pi^o$ decay) and radio emission from 
secondary electrons depends on
the properties of the magnetic field in the IGM.
Consequently 
limits on $\gamma$-rays from nearby clusters combined
with constraints from RM allow
for testing secondary models for radio halos (e.g., Ackermann et al 2010,
Donnert et al 2010a, Jeltema \& Profumo 2011 for recent attempts).

\noindent
A step in this direction can be obtained by 
{\it adding} the constraints given by the spatial
profile of the synchrotron brightness of radio halos.
It is well known that giant radio halos have flat brightness distributions
(eg Govoni et al 2001, Murgia et al 2009, Brown \& Rudnick 2011) 
implying that most of the emission is produced from the 
external regions where the magnetic field
is smaller. This immediately leads to a larger ratio gamma/radio 
emission in the case of radio halos with flatter radio profiles.

\noindent
Here we report on the case of Coma. 
We assume secondary models for the origin of the
radio halo and ``force'' the model
to reproduce the observed halo's radio profile and luminosity at 330 MHz
(see also Donnert et al 2010a).
This, combined with the FERMI limits, 
allows for obtaining corresponding (lower)
limits on the cluster's magnetic field.
In Fig.~2 we report a comparison between the magnetic field in the
cluster center $B_o$ ($B=B_o (n/n_o)^{\eta}$, $n$ the IGM density), derived 
from RM, and the minimum value of $B_o$ that is required
by secondary models to have $\gamma$-ray emission (still) consistent with 
the FERMI upper limits.
This starts putting tension on a secondary origin of the halo: 
the limits on $B_o$ imposed by FERMI (18 months
of observations, Ackermann et al.2010) are inconsistent with the values 
of $B_o$ derived (at 1, 2, 3$\sigma$ level) from RM.
Future FERMI data will be crucial as $\gamma$-ray upper
limits 50\% deeper
imply values of $B_o$ 1.5--2 times larger (depending on $\eta$).

\section{Conclusions}

Present observations put constraints on the nature of giant 
radio halos. Constraints come from the combination of (i) the spectral and
statistical properties of the population of radio halos, (ii) estimates
of $B$ from RM, and (iii) upper limits on $\gamma$-ray emission from 
radio-halo clusters.
The scenario for the origin of radio halos based on particle acceleration 
by merger-driven turbulence in galaxy clusters shows a good correspondence 
with the combination of these constraints.

\noindent
The physics of the interaction between turbulence and particles is however
very complex leaving many aspects of this scenario still unexplored.
We have discussed some open questions, e.g. {\it which is the fraction of the
energy of turbulence that goes into particle (re)acceleration in the IGM ?}
{\it How radio halos are generated and fade away in connection with
cluster mergers ?} {\it Why several merging clusters do not host
observable radio halos ?}

\noindent
We conclude that
observations with future radiotelescopes 
and deeper
constraints in the $\gamma$
-ray band will have the potential to test this 
model and (eventually) to further increase present difficulties with other
scenarii.

\section*{Acknowledgments} 

We thank partial support from PRIN INAF 2008/2009. We also thank the
SOC for organising such interesting conference.


\begin{thebibliography}{}

\bibitem{Ackermann2010} 
Ackermann M., et al. 2010, ApJ, 717, L71

\bibitem{Berezinsky1997}
Berezinsky V.S., Blasi P., Ptuskin V.S., 1997, ApJ, 487, 529

\bibitem{Blasi1999}
Blasi P., Colafrancesco S., 1999, APh, 12, 169

\bibitem{Blasi2007}
Blasi P., Gabici S., Brunetti G., 2007, IJMPA 22, 681

\bibitem{Bonafede2010}
Bonafede A., et al., 2010, A\&A, 513, 30

\bibitem{Bonafede2011}
Bonafede A., et al., 2011, arXiv:1103.0277

\bibitem{Brentjens2008}
Brentjens M.A., 2008, A\&A, 489, 69

\bibitem{Brown2011}
Brown S., Rudnick L., 2011, MNRAS, 412, 2

\bibitem{} Brunetti, G., 2003, ASP Conference Series, 301, 349

\bibitem{Brunetti2001}
Brunetti, G., et al., 2001, MNRAS, 320, 365

\bibitem{} Brunetti G., et al., 2004, MNRAS 350, 1174

\bibitem{} Brunetti G., Lazarian A., 2007, MNRAS 378, 245

\bibitem{Brunetti2008}
Brunetti, G., et al., 2008, Nature, 455, 944

\bibitem{Brunetti2009}
Brunetti, G., et al., 2009, A\&A, 507, 661

\bibitem{} Brunetti G., Lazarian A., 2011a, MNRAS, 412, 817

\bibitem{} Brunetti G., Lazarian A., 2011b, MNRAS, 410, 127

\bibitem{} Buote, D.~A., 2001, ApJ, 553, L15 

\bibitem{} Cassano R., Brunetti G., 2005, MNRAS 357, 1313

\bibitem{} Cassano, R., Brunetti, 
G., Setti, G., 2006, MNRAS, 369, 1577 

\bibitem{} Cassano, R., et al., 2008, A\&A, 480, 687 

\bibitem{} Cassano, R., et al., 2010a, ApJL, 721, L82

\bibitem{} Cassano, R., et al., 2010b, A\&A, 509, 68

\bibitem{} Cho, J., \& Lazarian, A., 2006, ApJ, 638, 811

\bibitem{} Colafrancesco, S., Profumo, S., Ullio, P., 2006, A\&A, 455, 21 

\bibitem{} Dallacasa D., et al., 2009, ApJ, 699, 1288

\bibitem{} Deiss, B.~M., et al., 1997, A\&A, 321, 55 

\bibitem{} Dolag, K., et al., 2005, MNRAS, 364, 753

\bibitem{Donnert2010a}
Donnert J., et al., 2010a, MNRAS, 401, 47

\bibitem{Donnert2010b} 
Donnert J., et al., 2010b, MNRAS, 407, 1565

\bibitem{} Ferrari, C. et al. 2008, SSRv 134, 93

\bibitem{Fujita2003} Fujita Y., Takizawa M., Sarazin C.L., 
2003, ApJ, 584, 190

\bibitem{} Giovannini, G., et al., 1993, ApJ, 406, 399 

\bibitem{Giovannini1999} 
Giovannini, G., Tordi, M., Feretti, L., 1999, NewA, 4, 141

\bibitem{} Giovannini, G., et al., 2009, A\&A, 507, 1257 

\bibitem{} Govoni F., et al., 2001, A\&A 369, 441

\bibitem{} Govoni, F., et al., 2004, ApJ, 605, 695 

\bibitem{Jaffe1977} Jaffe W.J., 1977, ApJ, 212, 1

\bibitem{Jeltema2011} Jeltema T.E., Profumo S., 2011, ApJ, 728, 53

\bibitem{} Kempner, J.~C., Sarazin, C.~L., 2001, ApJ, 548, 639 

\bibitem{Keshet2010} Keshet R., Loeb A., 2010, ApJ, 722, 737

\bibitem{} Iapichino, L., Niemeyer, J.~C., 2008, MNRAS, 388, 1089 

\bibitem{} Lazarian A., 2006, ApJ 645, L25

\bibitem{Macario2010} Macario, G., et al., 2010, A\&A, 517, A43

\bibitem{} Miniati, F.\ 2003, MNRAS, 342, 1009

\bibitem{Murgia2009} Murgia, M., et al., 2009, A\&A, 499, 679 

\bibitem{} Ohno, H., Takizawa, M., Shibata, S., 2002, ApJ, 577, 658 

\bibitem{} Paul, S., et al., 2011, ApJ, 726, 17

\bibitem{Petrosian2001} Petrosian V., 2001, ApJ, 557, 560 

\bibitem{Petrosian2008} Petrosian V., East W.E., 2008, ApJ, 682, 175

\bibitem{} Petrosian, V., Bykov, A.~M., 2008, SSRv, 134, 207

\bibitem{} Pizzo, R., 2010, PhD Thesis, Groningen University

\bibitem{Pfrommer2004} Pfrommer C., Ensslin T.A., 2004, MNRAS, 352, 76

\bibitem{} Rottgering, H.~J.~A., 2010, ISKAF2010 Science Meeting, Published
online at http://pos.sissa.it/cgi-bin/reader/conf.cgi?confid=112, p.50

\bibitem{} Russell, H. R., et al., 2011, arXiv: 1105.0435

\bibitem{Schekochihin2005} Schekochihin A.A., et al., 2005, ApJ, 629, 139

\bibitem{Schlickeiser1987} Schlickeiser R., et al., 1987, A\&A, 182, 21

\bibitem{} Thierbach M., Klein U., Wielebinski R., 2003, A\&A 397, 53

\bibitem{} Vazza, F., et al., 2011, A\&A, 529, A17 

\bibitem{} Venturi, T., 2011, arXiv:1102.1572

\bibitem{} Venturi, T., et al., 2007, A\&A 463, 937

\bibitem{} Venturi, T., et al., 2008, A\&A, 484, 327

\bibitem{Voelk1996} V\"{o}lk H.J., et al., 1996, SSRv, 75, 279

\bibitem{} Yan H., Lazarian A., 2004, ApJ 614, 757

\bibitem{} Willson, M.~A.~G., 1970, MNRAS, 151, 1 

\end{thebibliography}
\end{document}